# Error Rates of the Maximum-Likelihood Detector for Arbitrary Constellations: Convex/Concave Behavior and Applications[1]

Sergey Loyka, Victoria Kostina, Francois Gagnon

*Abstract* – Motivated by a recent surge of interest in convex optimization techniques, convexity/concavity properties of error rates of the maximum likelihood detector operating in the AWGN channel are studied and extended to frequency-flat slow-fading channels. Generic conditions are identified under which the symbol error rate (SER) is convex/concave for arbitrary multi-dimensional constellations. In particular, the SER is convex in SNR for any one- and two-dimensional constellation, and also in higher dimensions at high SNR. Pairwise error probability and bit error rate are shown to be convex at high SNR, for arbitrary constellations and bit mapping. Universal bounds for the SER $1^{st}$ and $2^{nd}$ derivatives are obtained, which hold for arbitrary constellations and are tight for some of them. Applications of the results are discussed, which include optimum power allocation in spatial multiplexing systems, optimum power/time sharing to decrease or increase (jamming problem) error rate, an implication for fading channels ("*fading is never good in low dimensions*") and optimization of a unitary-precoded OFDM system. For example, the error rate bounds of a unitary-precoded OFDM system with QPSK modulation, which reveal the best and worst precoding, are extended to arbitrary constellations, which may also include coding. The reported results also apply to the interference channel under Gaussian approximation, to the bit error rate when it can be expressed or approximated as a non-negative linear combination of individual symbol error rates, and to coded systems.

# I. INTRODUCTION

Optimization problems of various kinds simplify significantly if the goal and constraint functions involved are convex. Indeed, a convex optimization problem has a unique global solution, which can be found either analytically or, with a reasonable effort, by several efficient numerical methods (e.g. interior point methods); its numerical complexity grows only moderately with the problem dimensionality and required accuracy; convergence rates and required step size can be estimated in advance; there are powerful analytical tools that can be used to attack a problem and that provide insights into such problems even if solutions, either analytical or numerical, are not found yet [1][2]. Convex problems are almost as easy as liner ones. Contrary to this, not only generic nonlinear optimization problems do not possess these features, they are not solvable numerically, i.e. their complexity grows prohibitively fast with problem dimensionality and required accuracy [2]. Thus, there is a great advantage in formulating or at least in approximating an optimization problem as a convex one.

In the world of digital communications, one of the major performance measures is either symbol or bit error rate (SER or BER). Consequently, when an optimization of a communication system is performed, either SER or BER often appears as goal or constraint functions. Examples include optimum power/rate allocation in spatial multiplexing systems (BLAST) [3]-[6], optimum power/time sharing for a transmitter and a jammer [7], rate allocation or precoding in multicarrier (OFDM) systems [8][9], optimum equalization [10], optimum multiuser detection [11][21], and joint Tx-Rx beamforming (precoding-decoding) in MIMO systems [12]. Symbol and bit error rates of the maximum likelihood (ML) detector have been extensively studied and a large number of exact or approximate analytical results are available for various modulation formats, for both non-fading and fading AWGN channels [13]-[17]. One- and two-dimensional (1-D and 2-D) constellations have been studied in greater depth [30]-[34], and exact analytical expressions for SER and BER of arbitrary PAM and QAM [18] as well as efficient numerical techniques for arbitrary 2-D constellations [19] are available. A generic parameterization of error rates in fading channels at high SNR via diversity and coding gains have been presented in [20].

While the error rates themselves have been a subject of intensive studies, their convexity/concavity properties, which are so important for optimization, have not been studied in depth; the results in this area are very limited. Convexity/concavity properties of the Q-function, $Q(x) = \sqrt{2\pi}^{-1} \int_x^\infty e^{-t^2/2} dt$, are well known: $Q(x)$ and $Q(\sqrt{x})$ are convex for $x \geq 0$ [21] (convexity in amplitude and SNR respectively), from which it follows that any combination of the form $\sum_i \alpha_i Q(\beta_i x)$ or $\sum_i \alpha_i Q(\sqrt{\beta_i x})$, where $\alpha_i, \beta_i \geq 0$ are constants, is also convex. The last combination approximates well many modulation formats, and the Q-function itself gives the error rate of coherent BPSK and QPSK, and also approximates the error rates of several modulation formats (e.g. using the nearest neighbor argument) [13][15]-[17][30][31]. Non-coherent BPSK and QPSK error rates are expressed via $\alpha \exp(-\beta x)$, which is also convex; the same function or its combinations of the form $\sum_i \alpha_i \exp(-\beta_i x)$, which are also convex, approximate well a few other modulation formats and also serve as an upper bound (Chernoff or union bounds) for many more, including those in MIMO systems[2]. Most known closed-from expressions for error rates (e.g. in [13]-[17]) can be verified, by differentiation, to be convex. Little is known besides that[3]. Is the SER/BER convex for all 1-D or 2-D constellations of arbitrary shape? What about arbitrary multi-dimensional constellations? Under what conditions? In what variables, i.e. SNR, amplitude, 1/SNR (e.g. noise power), 1/amplitude (e.g. noise amplitude)? What about fading channels in general? To make use of all the important features and powerful algorithms of convex optimization in digital communications (see e.g. [43]) on a rigorous footing, these questions have to be answered.

The present paper aims at answering these questions in a systematic way by developing a geometric method of the SER representation specifically tailored for its convexity analysis. Convexity properties of error rate for binary modulations in terms of SNR and noise power have been studied in detail in [7]. Here, we generalize the results in [7] to the constellations of arbitrary shape, order and dimensionality operating with the maximum likelihood detector in the AWGN channel, for both non-fading and frequency-flat slowly-fading channels. While most of our results are derived for the SER, some of them also apply to the BER when the later can be expressed or approximated as a non-negative linear combination of corresponding SER expressions; see [18][33] for details on such approximations. With Gray encoding and when nearest neighbor errors dominate, the BER can be approximated as $\frac{1}{\log_2 M}$ SER [16][17], where $M$ is the number of constellation points, which obviously inherits the convexity property from the SER.

The paper is organized as follows. Section II introduces the system model. We consider the maximum

---

[2] Unfortunately, nothing can be said about convexity using this approach when some $\alpha_i$ are negative. In this case, however, approximations are often obtained that include only positive terms (see e.g. [13]-[17][32][34]) to which this approach applies. Furthermore, the BER can be presented as a positive linear combination of pairwise error probabilities [33], which we exploit in Corollary 4.1.

[3] After this paper has been submitted, Conti et al [29] has presented a log-concavity property of the SER in SNR [dB] for the uniform square-grid constellations.

18-Nov-09                                             IT Trans.: 2nd Revision                                            3(30)

likelihood detector that operates symbol-by-symbol (no memory) in the AWGN channel, which is later extended to frequency-flat slow-fading channels with a generic SNR distribution (e.g. not limited to Rayleigh fading); no any specific assumptions about constellation geometry, order or dimensionality are made.

Section III analyzes the convexity/concavity of error rates in SNR, amplitude and noise power using a systematic geometric method that does not rely on any constellation properties or approximations but rather exploits the spherical symmetry of the Gaussian noise distribution. The SER is shown to be convex in SNR for arbitrary 1-D and 2-D constellations. For 3-D and higher-D constellations, the SER is convex at large SNR, concave at small SNR, and has an odd number of inflection points in-between. It is shown in Section VII that this non-convexity can be used to reduce the SER of higher-D constellations via a time/power sharing algorithm under the fixed average power constraint, which is impossible for any 1- and 2-D constellations[4]. Using the same approach, we show that the pairwise error probability (PEP) and, thus, the BER are always convex at high SNR, for any bit mapping. However, unlike the SER, the PEP can be concave at low SNR, even for 1-D and 2-D constellations. In the case of log-concave but otherwise arbitrary noise density (e.g. Gaussian, Laplacian, exponential), the probability of correct decision is shown to be log-concave, which suffices for optimization problems that maximize/minimize this probability.

The study of the convexity property of SER in the noise power is motivated by the jammer optimization problem [7]. The SER is shown to be concave in the noise power at low SNR (large noise), convex at high SNR (small noise) and has an odd number of inflection points in-between. This result is used in Section VII to find the optimum power/time sharing of the jammer (the noise source) to increase the SER, based on the technique developed in [7] for binary modulations. This result can also be applied to the SER as a function of the mean square error (MSE), as in the precoder or equalizer design problems [9][12][21][26]-[28].

Section IV presents a number of lower and upper bounds on first and second derivatives of the SER in SNR and noise power, which hold for arbitrary constellations and depend only on their dimensionality. Such bounds are important as the derivatives play a prominent role in the design and analysis of numerical optimization algorithms for a number of reasons [1]: to analyze the convergence conditions and rates, to determine the step size of gradient methods and to assess sub-optimality of various solutions, which is further used as a stopping criteria. The derivatives in noise power find applications in the jammer optimization problem (see Sections VII-B,C and [7]).

---

[4] This impossibility for binary modulations has been first pointed out in [7], and is extended here to all 1 and 2-D constellations in the AWGN channel.

Section V deals with a frequency-flat slowly fading channel. It is shown that the average SER is convex provided that the instantaneous SER is convex and a mild condition on the distribution of the instantaneous SNR holds.

Since the Q-function finds wide applications in communication/information theory, we derive in Section VI a number of new convexity/concavity properties of this function, which complement the known ones (see [21] for an extensive list of such properties).

Section VII deals with several applications of the convexity results. It is demonstrated that the optimum power allocation in the V-BLAST algorithm with the zero-forcing (ZF) successive interference cancellation (SIC) interface has a unique global solution for all 1- and 2-D constellations, but not necessarily for higher-D ones, both in non-fading and fading channels, which extends the corresponding results in [3]-[5] and also applies to power allocation in OFDM systems [41]. The optimum and simple sub-optimum power/time sharing strategies of a jammer are developed to maximize the SER, which extend the corresponding results in [7] to arbitrary multi-dimensional constellations in the AWGN channel. It is shown that there exists no fading distribution that can reduce the SER (compared to the non-fading channel) of arbitrary 1 and 2-D constellations, i.e. "*fading is never good in low dimensions*". This does not hold for higher-D constellations. Finally, known bounds on the error rate of unitary-precoded OFDM system with QPSK modulation and optimum precoding [9] are extended to arbitrary constellations (possibly with coding), which reveals the best and worst transmission strategies.

While we do not consider explicitly interference (e.g. from multiple users), the results above also extend to the case of interference channel when it can be modeled as Gaussian noise, which is a popular approach in the literature (see e.g. [9][21][28]); in such a case, the noise power becomes the noise plus interference power, and the SNR is changed to SNIR, and all the results above hold. The reported results also apply to the BER when it can be expressed as a non-negative linear combination of individual symbol error rates, and also to modulation with coding, by considering codewords as symbols of an extended multi-dimensional constellation. While error rate performance analysis of coded systems is a formidable analytical task so that bounding is the only solution in most cases [44], our approach allows one to establish the convexity properties of error rates of such systems without resorting to error rate results or bounds. This opens up an opportunity to use powerful convex optimization techniques for the design and optimization of coded systems on a rigorous footing.

## II. SYSTEM MODEL

The standard baseband discrete-time system model of an AWGN channel, which includes matched filtering and sampling, is adopted here,

$$\mathbf{r} = \mathbf{s} + \boldsymbol{\xi} \tag{1}$$

where $\mathbf{s}$ and $\mathbf{r}$ are $n$-dimensional vectors representing the Tx and Rx symbols respectively, $\mathbf{s} \in \{\mathbf{s}_1, \mathbf{s}_2, ..., \mathbf{s}_M\}$, $\{\mathbf{s}_1, \mathbf{s}_2, ..., \mathbf{s}_M\}$ is a set of $M$ constellation points, $\boldsymbol{\xi}$ is the additive white Gaussian noise (AWGN), $\boldsymbol{\xi} \sim \mathcal{N}(\mathbf{0}, \sigma_0^2 \mathbf{I})$, whose probability density function (PDF) is

$$p_\xi(\mathbf{x}) = \left(\frac{1}{2\pi\sigma_0^2}\right)^{\frac{n}{2}} e^{-\frac{|\mathbf{x}|^2}{2\sigma_0^2}} \tag{2}$$

where $\sigma_0^2$ is the noise variance per dimension, and $n$ is the constellation dimensionality; lower case bold letters denote vectors, bold capitals denote matrices, $x_i$ denotes $i$-th component of $\mathbf{x}$, $|\mathbf{x}|$ denotes $L_2$ norm of $\mathbf{x}$, $|\mathbf{x}| = \sqrt{\mathbf{x}^T \mathbf{x}}$, where the superscript T denotes transpose, $\mathbf{x}_i$ denotes $i$-th vector. Frequency-flat slow-fading channels will be considered as well. The average (over the constellation points) SNR is defined as $\gamma = 1/\sigma_0^2$, which implies the appropriate normalization, $\frac{1}{M}\sum_{i=1}^{M}|\mathbf{s}_i|^2 = 1$.

Consider the maximum likelihood detector, which is equivalent to the minimum distance one in the AWGN channel [13]-[17],

$$\hat{\mathbf{s}} = \arg\min_{\mathbf{s}_i} |\mathbf{r} - \mathbf{s}_i| \tag{3}$$

The probability of symbol error $P_{ei}$ given than $\mathbf{s} = \mathbf{s}_i$ was transmitted is

$$P_{ei} = \Pr\left[\hat{\mathbf{s}} \neq \mathbf{s}_i \mid \mathbf{s} = \mathbf{s}_i\right] = 1 - P_{ci} \tag{4}$$

where $P_{ci}$ is the probability of correct decision. The SER averaged over all constellation points is

$$P_e = \sum_{i=1}^{M} P_{ei} \Pr[\mathbf{s} = \mathbf{s}_i] = 1 - P_c \tag{5}$$

$P_{ci}$ can be expressed as [15][17]

$$P_{ci} = \int_{\Omega_i} p_\xi(\mathbf{x}) d\mathbf{x} \tag{6}$$

where $\Omega_i$ is the decision region (Voronoi region), and $\mathbf{s}_i$ corresponds to $\mathbf{x} = \mathbf{0}$, i.e. the origin is shifted for convenience to the constellation point $\mathbf{s}_i$. $\Omega_i$ can be expressed as a convex polyhedron [1],

$$\Omega_i = \{\mathbf{x}: \mathbf{A}\mathbf{x} \leq \mathbf{b}\}, \quad \mathbf{a}_j^T = \frac{(\mathbf{s}_j - \mathbf{s}_i)}{|\mathbf{s}_j - \mathbf{s}_i|}, \quad b_j = \frac{1}{2}|\mathbf{s}_j - \mathbf{s}_i| \tag{7}$$

where $\mathbf{a}_j^T$ denotes j-th row of $\mathbf{A}$, and the inequality in (7) is applied component-wise.

## III. CONVEXITY OF ERROR RATES IN SNR, AMPLITUDE AND NOISE POWER

Since many optimization problems in digital communications use error rates as either the goal or constraint functions (see [3]-[12] for examples), and since the optimization is often carried out under various power constraints, we begin the analysis by studying the convexity properties of the SER in terms of the SNR (which is equivalent to the signal power or energy under fixed noise) for arbitrary constellations.

### A. Convexity of the SER in the SNR and signal amplitude

**Theorem 1:** $P_e(P_c)$ is a convex (concave) function of the SNR $\gamma$ for any constellation of dimensionality $n \leq 2$,

$$\frac{d^2 P_e}{d\gamma^2} = P''_{e|\gamma} > 0, \quad P''_{c|\gamma} < 0 \tag{8}$$

**Proof:** see Appendix A.

Theorem 1 covers, as special cases, such popular constellations as BPSK, BFSK, QPSK, QAM, M-PSK, OOK, whose error rate convexity can also be verified directly by differentiation of known error rate expressions. The convexity property of the SER also extends to the BER, when the later can be expressed or approximated as a linear combination of error rates of individual symbols of the constellation, i.e. when $\text{BER} = \sum_i \alpha_i P_{ei}(\beta_i \gamma)$, $\alpha_i, \beta_i \geq 0$ (this holds for a number of BER expressions or their approximations for PAM and QAM constellation with Gray encoding [15][34][30]).

While 1- and 2-D constellation always have convex symbol error rates, higher-D constellations exhibit more complicated behavior, as shown below.

**Theorem 2:** For constellations of dimensionality $n > 2$, $P_{ei}$ ($P_{ci}$) has the following convexity properties,

2.1. $P_{ei}$ ($P_{ci}$) is convex (concave) in the large SNR mode,

$$\gamma \geq \frac{n + \sqrt{2n}}{d^2_{\min,i}} \tag{9}$$

where $d_{\min,i} = \min_j (b_j)$ is the minimum distance from the origin to the boundary of decision region $\Omega_i$ (see Appendix A).

2.2. $P_{ei}$ ($P_{ci}$) is concave (convex) in the small SNR mode,

$$\gamma \leq \frac{n - \sqrt{2n}}{d^2_{\max,i}} \tag{10}$$

where $d_{max,i}$ is the maximum distance from the origin to the boundary of $\Omega_i$ [5].

2.3. There are an odd number of inflection points, $P_{ci|\gamma}'' = P_{ei|\gamma}'' = 0$, in the intermediate SNR mode,

$$\frac{n-\sqrt{2n}}{d_{max,i}^2} \leq \gamma \leq \frac{n+\sqrt{2n}}{d_{min,i}^2} \quad (11)$$

**Proof:** see Appendix A.

While Theorem 2 applies to conditional error rates $P_{ei}$, similar result also holds for the error rate $P_e$ averaged over the constellation.

**Corollary 2.1**: Using the fact that non-negative weighted sum of convex (concave) functions is also convex (concave), the results in Theorem 2 extend directly to $P_e$ ($P_c$) via the substitutions $d_{max,i} \to d_{max} = \max_i \{d_{max,i}\}$ and $d_{min,i} \to d_{min} = \min_i \{d_{min,i}\}$ in (9)-(11),

1. $P_e$ ($P_c$) is convex (concave) in the large SNR mode, $\gamma \geq (n+\sqrt{2n})/d_{min}^2$.

2. It is concave (convex) in the small SNR mode, $\gamma \leq (n-\sqrt{2n})/d_{max}^2$.

3. There are an odd number of inflection points, $P_{c|\gamma}'' = P_{e|\gamma}'' = 0$, in the intermediate SNR mode, $(n-\sqrt{2n})/d_{max}^2 \leq \gamma \leq (n+\sqrt{2n})/d_{min}^2$. ∎

Theorem 2 indicates that the constellation dimensionality plays an important role for concavity/convexity properties. Below we present a result which is independent of the dimensionality and holds for a wide class of channels.

**Theorem 3:** $P_{ci}$ is log-concave in SNR for arbitrary constellation, arbitrary $n$ and any log-concave noise density (i.e. Gaussian, Laplacian, exponential, see [35] for an extensive list of such densities and their properties).

**Proof**: via the integration theorem for log-concave functions [1, p.106][35].

Unfortunately, in the general case log-concavity does not extend to $P_c$ since the sum of log-concave functions is not necessarily log-concave. However, in some special cases it does.

**Corollary 3.1:** $P_c$ is log-concave under the conditions of Theorem 3 for a symmetric constellation, i.e. for $P_e = P_{e1} = P_{e2} = ... = P_{eM}$ (e.g. the uniform signal sets [17]).

**Proof:** immediate from Theorem 3 since $P_c = P_{ci}$.

We note that log-concavity is a "weaker" property than concavity as the latter does not follow from the former [1].

---

[5] It should be noted that the small SNR regions in (10) do not exist if $d_{max} = \infty$, i.e. unbounded $\Omega_i$. This, however, does not imply that $P_{ei}$ is always convex, because of the intermediate SNR region in (11), where $P_{ei}$ may be concave over a certain interval. Also note that the conditions for convexity/concavity in (9), (10) are sufficient but not necessary (e.g $P_{ei}$ may also be convex outside of the interval in (9)).

Yet, it is useful for many optimization problems, which can be reformulated in terms of $\log P_c$ and thus can be treated as convex optimization problems[6], with all the advantages mentioned above.

In some cases (e.g. in inter-symbol interference analysis, equalizer design/optimization etc. [38]-[40][27][28]), error rate is considered as a function of the signal amplitude $A = \sqrt{\gamma}$ rather than power or SNR, so its convexity in $A$ is of interest. This can be studied using the same geometric approach as in Theorem 1 and 2, which is summarized below:

- The SER as a function $F_i(A) = P_{ei}(A^2)$ of the amplitude $A$ is always convex for $n = 1$,
- for $n \geq 2$, it is convex at high SNR $\gamma = A^2 \geq \alpha_1 / d_{\min,i}^2$ and concave at low SNR $A^2 \leq \alpha_2 / d_{\max,i}^2$, where,

$$\alpha_1 = \frac{1}{2}\left(2n + 1 + \sqrt{8n+1}\right), \; \alpha_2 = \frac{1}{2}\left(2n + 1 - \sqrt{8n+1}\right),$$

  and there is an odd number of inflection points in –between;
- the same applies to $P_e(A^2)$ via $d_{\min,i} \to d_{\min}$, $d_{\max,i} \to d_{\max}$.

Note that convexity in the amplitude is a stronger property than convexity in the SNR (power) as the latter follows from the former (via the composition rule) but not the other way around: while the SER is always convex in the SNR for $n = 2$, it can be concave in $A$.

*B. Convexity of the PEP and BER in the SNR*

In many cases, it is a pairwise error probability (PEP), i.e. a probability $\Pr\{\mathbf{s}_i \to \mathbf{s}_j\} = \Pr[\hat{\mathbf{s}} = \mathbf{s}_j | \mathbf{s} = \mathbf{s}_i]$ to decide in favor of $\mathbf{s}_j$ given that $\mathbf{s}_i$, $i \neq j$, was transmitted[7], that is a key point in the analysis (e.g. for constructing a union bound and other performance metrics, see e.g. [13][15]-[17][37]). Its convexity property can be established in a way similar to Theorem 1, 2.

**Theorem 4**:

a) the pairwise error probability $\Pr\{\mathbf{s}_i \to \mathbf{s}_j\}$ is a convex function of the SNR at the high SNR region, $\gamma \geq (n + \sqrt{2n}) / d_{\min,i}^2$, for any $n$,

b) it is concave for $n = 1, 2$ at the low SNR region, $\gamma \leq (n + \sqrt{2n}) / (d_{ij} + d_{\max,j})^2$, where $d_{ij} = |\mathbf{s}_i - \mathbf{s}_j|$ is the distance between $\mathbf{s}_i$ and $\mathbf{s}_j$,

---

[6] maximizing (minimizing) $P_c$ is equivalent to maximizing (minimizing) $\log P_c$ since $\log(\;)$ is a monotonically increasing function.

[7] Our definition of the PEP is slightly different from the conventional one (which is a probability that $\mathbf{r}$ is closer to $\mathbf{s}_i$ than to $\mathbf{s}_j$ [16][17]) so that our PEP $\neq Q(|\mathbf{s}_i - \mathbf{s}_j| / (2\sigma_0))$, which has important advantages: $P_{ei} = \sum_j \Pr\{\mathbf{s}_i \to \mathbf{s}_j\}$ and the BER is expressed as in (12), which is impossible with the conventional definition.

c) it is convex for $n > 2$ at the low SNR region, $\gamma \leq (n - \sqrt{2n})/(d_{ij} + d_{max,j})^2$.

**Proof:** along the same lines as that of Theorem 1 and 2.

We note that Theorem 4(a) is stronger than Theorem 1 (at the high SNR region) since the latter follows from the former but the opposite is not always true (as the other SNR range in Theorem 1 indicates). Unlike the SER, the pairwise error probability can be concave at low SNR even for $n = 1, 2$.

It follows from Theorem 4 that the BER is also convex in the high SNR region.

**Corollary 4.1:** Under the conditions of Theorem 4(a) with the substitution $d_{min,i} \to d_{min} = \min_i \{d_{min,i}\}$,

$$\gamma \geq (n + \sqrt{2n})/d_{min}^2,$$

the pairwise error probabilities $\Pr\{s_i \to s_j\} \ \forall i, j$, and also the BER (for any bit mapping) are convex functions of the SNR.

**Proof:** using the relationship between the BER and the pairwise error probabilities [33],

$$\text{BER} = \sum_{i=1}^{M} \sum_{j \neq i} \frac{h_{ij}}{\log_2 M} \Pr\{s = s_i\} \Pr\{s_i \to s_j\} \tag{12}$$

where $h_{ij}$ is the Hamming distance between binary sequences representing $s_i$ and $s_j$, and observing that a positive linear combination of convex functions is convex. *Q.E.D.*

Remarkably, the high-SNR conditions in Corollary 4.1 and 2.1 are the same, i.e. not only the SER, but also the PEP and the BER are convex in this high SNR range. In some cases (Gray encoding and when nearest neighbor errors dominate), the BER is approximated as $\frac{1}{\log_2 M}$ SER [16][17], so that it inherits the convexity properties from Theorems 1, 2.

*C. Convexity of the SER in Noise Power*

Following the same geometric approach as in Section III-A, we study below the convexity properties of $P_{ei}$ ($P_{ci}$) as functions of the noise power, which has applications in the jammer optimization problem (see Sections VII-B,C) and also in the optimization problems which express the error rate as a function of the MSE (e.g. equalizer or precoder design [9][12], see Section VII-E, where the MSE is considered as a part of the noise).

**Theorem 5:** $P_{ei}$ has the following convexity properties in the noise power $P_N = \sigma_0^2$, for any n,

4.1. $P_{ei}$ is concave in the large noise mode (low SNR),

$$P_N \geq \frac{d_{max,i}^2}{n + 2 - \sqrt{2(n+2)}} \tag{13}$$

4.2. $P_{ei}$ is convex in the small noise mode (high SNR),

$$P_N \le \frac{d_{\min,i}^2}{n+2+\sqrt{2(n+2)}} \tag{14}$$

4.3. There are an odd number of inflection points for intermediate noise power,

$$\frac{d_{\min,i}^2}{n+2+\sqrt{2(n+2)}} \le P_N \le \frac{d_{\max,i}^2}{n+2-\sqrt{2(n+2)}} \tag{15}$$

**Proof:** see Appendix A.

**Corollary 5.1**: The results in Theorem 5 extend directly to $P_e$ ($P_c$) by the substitutions $d_{\max,i} \to d_{\max}$ and $d_{\min,i} \to d_{\min}$ in (13)-(15).

## IV. Universal Bounds on SER Derivatives in SNR and Noise Power

Here we explore some generic properties of the SER derivatives in SNR and noise power, which hold for arbitrary constellations, based on the results in Sections III. Such derivatives play an important role in the design and analysis of optimization algorithms for several reasons: to analyze the convergence conditions and rate, to assess sub-optimality of found solutions and thus to develop a stopping criteria, and others (see chapters 9-11 in [1] for more details). Since the bounds developed below hold for arbitrary constellations, they can be used in optimization algorithms applicable to a wide class of problems.

**Theorem 6:** The first derivative in SNR $P'_{e|\gamma}$ (and also $P'_{ei|\gamma}$) is bounded, for arbitrary constellation, as follows,

$$-\frac{c_n}{\gamma} \le P'_{e|\gamma} \le 0 \tag{16}$$

where $c_n = (n/2)^{n/2} e^{-n/2}/\Gamma(n/2)$, where $\Gamma(\ )$ is the gamma-function (see (A14)) [23].

**Proof:** see Appendix A.

It should be noted that the bounds depend only on the SNR and constellation dimensionality, not on constellation geometry or order. They also apply to $P'_{ei|\gamma}$. Note that $|P'_{e|\gamma}|$ decreases with SNR at least as $1/\gamma$.

*Example:* for arbitrary constellation geometry and order, the SER 1$^{st}$ derivative in the SNR is bounded as follows:

$$-\frac{1}{\sqrt{2\pi e \gamma}} \le P'_{e|\gamma} \le 0, \quad n=1 \tag{17}$$

$$-\frac{1}{e\gamma} \le P'_{e|\gamma} \le 0, \quad n=2 \tag{18}$$

When dimensionality is large ($n \gg 1$), $c_n \approx \sqrt{\frac{n}{4\pi}}$ and the upper bound on $|P'_{e|\gamma}|$ increases with $n$, i.e.

18-Nov-09             IT Trans.: 2$^{nd}$ Revision             11(30)

higher-dimensional constellations (which may also include coding) have potential for faster decrease of error rates with the SNR.

**Corollary 6.1:** When the lower bound in (16) is applied to $P'_{ei|\gamma}$, it is achieved for the spherical decision region, $\Omega_i = C^+ = \{\mathbf{x}: |\mathbf{x}|^2 \leq n/\gamma\}$, of the radius $\sqrt{n/\gamma} = \sqrt{n\sigma_0^2}$.

**Proof:** immediate from the proof of Theorem 6.

While the spherical decision region is not often encountered in uncoded systems, it has a number of remarkable properties: it is the best possible decision region in the sense that it minimizes the error probability for the symbol it represents [15]; it is a decision region for some non-coherent constellations [17]; and it enters intimately into the channel coding theorem [15][42] (via the sphere hardening and packing arguments) so that capacity-approaching codes should have near-spherical decision regions and the bounds above become tight. We note however that they can be tight only for a specific SNR for non-adaptive systems (fixed constellation) as the sphere radius in Corollary 6.1. depends on the SNR while the constellation geometry does not.[8] The SNR at which the bound is achieved satisfies to $1/\sigma_0^2 = n/R^2$ so that the "effective" SNR for this symbol is $\gamma_{eff} = R^2/\sigma_0^2 = n$.

**Corollary 6.2:** The asymptotic behavior of $P'_{ei|\gamma}$ and $P'_{ci|\gamma}$, which also applies to $P'_{e|\gamma}$ and $P'_{c|\gamma}$, is as follows

$$\lim_{\gamma \to \infty} P'_{ei|\gamma} = \lim_{\gamma \to \infty} P'_{ci|\gamma} = 0 \qquad (19)$$

and the convergence to the limit is uniform.

**Proof:** immediate from Theorem 6.

The intuition behind this result is that while $P'_{e|\gamma} < 0$ (error rate decreases with SNR), from Theorem 1 and 2 $P''_{e|\gamma} > 0$ at high SNR (convexity), so that eventually, as SNR increases, $P'_{e|\gamma}$ has to approach zero[9].

In a similar way, one can derive bounds on the second derivative of the SER.

**Theorem 7:** The second derivative in SNR $P''_{e|\gamma}$ (and also $P''_{ei|\gamma}$) is bounded, for arbitrary constellation, as follows,

$$\frac{B_l}{\gamma^2} \leq P''_{e|\gamma} \leq \frac{B_u}{\gamma^2} \qquad (20)$$

where $B_u = a_n (a_n)^{n/2} e^{-a_n} / \Gamma(n/2)$, $B_l = -(-b_n)_+ (b_n)^{n/2} e^{-b_n} / \Gamma(n/2)$, $a_n = (2+\sqrt{2n})/2$, $b_n = (2-\sqrt{2n})/2$, and $(x)_+ = x$ if $x \geq 0$ and 0 otherwise.

---

[8] This is the price to pay for the universal nature of the bound; naturally, when some specific information about the constellation is available, tighter bounds can be constructed.

[9] Note that when log-log scale is used ($\log P_e$ vs. $\gamma$[dB]) as in most SER/BER plots [13][15][17], (19) does not apply.

**Proof:** similar to that of Theorem 6, by observing that the lower and upper bounds, when applied to $P''_{ei|\gamma}$, correspond to the spherical decision regions of radii $R_l = \sqrt{(n-\sqrt{2n})_+/\gamma}$ and $R_u = \sqrt{(n+\sqrt{2n})/\gamma}$. Using (A13), the bounds can be immediately evaluated. Since the bounds do not depend on $\Omega_i$, they also apply to $P''_{e|\gamma}$. Q.E.D.

Note from (20) that $|P''_{e|\gamma}|$ decreases at least as $1/\gamma^2$, for any constellation.

*Example:* for any 2-D constellation in the AWGN channel, 2nd derivative of the SER is bounded as,

$$0 \le P''_{e|\gamma} \le \left(\frac{2}{e\gamma}\right)^2 \tag{21}$$

**Corollary 7.1:** the lower and upper bounds in (20) are achieved for the spherical decision regions of the radii $R_l$ and $R_u$.

**Proof:** immediate from the proof of Theorem 7.

The "effective" symbol SNR at which the bounds are achieved are $\gamma_l = R_l^2 \gamma = (n-\sqrt{2n})_+$ and $\gamma_u = R_u^2 \gamma = n+\sqrt{2n}$. As in the case of 1st derivative, spherical decision regions play here a prominent role.

**Corollary 7.2:** The asymptotic behavior of $P''_{ei|\gamma}$ and $P''_{ci|\gamma}$, which also applies to $P''_{e|\gamma}$ and $P''_{c|\gamma}$, is as follows

$$\lim_{\gamma \to \infty} P''_{ei|\gamma} = \lim_{\gamma \to \infty} P''_{ci|\gamma} = 0 \tag{22}$$

and the convergence to the limit is uniform.

**Proof:** immediate from Theorem 7.

The intuition behind this result is similar to that of Corollary 6.2: since $P'_{e|\gamma} < 0$ and $P''_{e|\gamma} > 0$ at high SNR, the second derivative has to approach zero to avoid positive first derivative.

**Corollary 7.3:** $P_{ei}$, $P_{ci}$ (and also $P_e$, $P_c$) and their first derivatives are continuous differentiable functions of the SNR.

**Proof:** immediate from Theorems 6 and 7.

Let us now explore properties of the SER derivatives in the noise power. These results parallel ones for the SNR derivatives and have similar proofs, which are omitted here for brevity.

**Theorem 8:** The first derivative in the noise power $P'_{e|P_N}$ is bounded, for arbitrary constellation, as follows,

$$0 \le P'_{e|P_N} \le \frac{c_n}{P_N} \tag{23}$$

**Corollary 8.1:** The upper bound in (23) is achieved for the spherical decision region of the radius $\sqrt{nP_N}$.

**Theorem 9:** The second derivative in the noise power $P''_{e|P_N}$ is bounded, for arbitrary constellation, as follows,

$$\frac{b_l}{P_N^2} \leq P''_{e|P_N} \leq \frac{b_u}{P_N^2} \tag{24}$$

where

$$b_u = \sqrt{\frac{n+2}{2}} \frac{(b_1)^{n/2} e^{-b_1}}{\Gamma(n/2)}, \quad b_l = -\sqrt{\frac{n+2}{2}} \frac{(b_2)^{n/2} e^{-b_2}}{\Gamma(n/2)}, \quad b_1 = \frac{n+2+\sqrt{2(n+2)}}{2}, \quad b_2 = \frac{n+2-\sqrt{2(n+2)}}{2}.$$

We note that $P'_{e|P_N}, |P''_{e|P_N}|$ decrease at least as $1/P_N$, $1/P_N^2$, respectively.

**Corollary 9.1**: the lower and upper bounds in (24) are achieved for the spherical decision regions of the radii

$$R_l = \sqrt{2b_2 P_N}, \quad R_u = \sqrt{2b_1 P_N} \tag{25}$$

with the effective SNRs $\gamma_l = R_l^2/P_N = n+2-\sqrt{2(n+2)}$ and $\gamma_u = R_u^2/P_N = n+2+\sqrt{2(n+2)}$.

**Corollary 9.2:** The asymptotic behavior of $P''_{ei|P_N}$ and $P''_{ci|P_N}$, which also applies to $P''_{e|P_N}$ and $P''_{c|P_N}$, is as follows

$$\lim_{P_N \to \infty} P''_{ei|P_N} = \lim_{P_N \to \infty} P''_{ci|P_N} = 0 \tag{26}$$

and the convergence to the limit is uniform.

**Corollary 9.3:** $P_{ei}$, $P_{ci}$ (and also $P_e$, $P_c$) and their first derivatives are continuous differentiable functions of the noise power.

Using the same method, the bounds for the 1st and 2nd derivatives, both in the SNR and the noise power, can also be extended to higher-order derivatives. The analysis, however, becomes more complicated.

## V. Convexity of Average SER in Fading Channels

Some of the convexity/concavity results above also apply to fading channels, which is explored in this section. We assume frequency-flat slow-fading channel.

**Theorem 10**: If the instantaneous SER $P_e$ is convex (concave) and the CDF of the instantaneous SNR $\gamma$ is a function of $\gamma/\gamma_0$ only,

$$CDF(\gamma) = F(\gamma/\gamma_0) \tag{27}$$

where $\gamma_0$ is the average SNR, then the average SER $\bar{P}_e$ is convex (concave) in $\gamma_0$.

**Proof:** consider the average SER,

$$\bar{P}_e = \int_0^\infty P_e(\gamma) PDF(\gamma) d\gamma = \int_0^\infty P_e(\gamma) dCDF(\gamma) = \int_0^\infty P_e(\gamma) dF(\gamma/\gamma_0)$$
$$= \int_0^\infty P_e(\gamma_0 t) dF(t) = \int_0^\infty P_e(\gamma_0 t) f(t) dt \tag{28}$$

where $PDF(\gamma) = dCDF(\gamma)/d\gamma$ is the PDF of $\gamma$, and $f(t) = dF(t)/dt \geq 0$ is the PDF of the normalized instantaneous SNR $t = \gamma/\gamma_0$. The convexity (concavity) of $\bar{P}_e$ follows from the last integral in (28), which is a non-negative weighted sum. *Q.E.D*.

In many cases, the large-SNR approximation of the error rate is used instead of the true SER, $\bar{P}_e \approx \text{contant}/\gamma_0^k$ [4][14][20]. It is straightforward to verify that this is also a convex function.

The equivalent to condition (27) is that the PDF of $\gamma$ can be expressed as $PDF(\gamma) = g(\gamma/\gamma_0)/\gamma_0$. The condition is not too restrictive as many popular fading channel models satisfy it, which includes Rayleigh fading channel (also with the maximum-ratio combining), Rice and Nakagami fading channels. However, some channels do not satisfy (27), which includes the log-normal and composite fading channels[10].

It should also be pointed out that Theorem 3 does not extend to fading channels in general, since the sum (or integral) of log-concave functions is not necessarily log-concave.

## VI. CONVEXITY PROPERTIES OF $Q(x)$

Since the $Q$ function $Q(x) = \frac{1}{\sqrt{2\pi}} \int_x^\infty e^{-t^2/2} dt$ finds extensive applications in communication/information theory, including many error rate expressions (it is the error rate of a binary modulation and many higher-order ones and their approximations and bounds, e.g. union bounds, include $Q(x)$ as a building block), we briefly summarize its convexity/concavity properties, which serves as a complement of the extensive list of its properties in [21]. A number of such convexity/concavity properties are well-known,

- $Q(x)$ is convex for $x \geq 0$ (convexity in amplitude) [21].
- $Q(\sqrt{x})$ is convex for $x \geq 0$ (convexity in power or SNR) [21].
- Linear combinations $\sum_i \alpha_i Q(\beta_i x)$ and $\sum_i \alpha_i Q(\sqrt{\beta_i x})$, where $\alpha_i, \beta_i \geq 0$ are constants, are also convex, which follows from the 1$^{st}$ two properties.
- $Q(1/\sqrt{x})$ is convex for $0 < x \leq 3$ and concave for $x > 3$ (convexity/concavity in noise power or MSE) [9]. Convexity/concavity of corresponding linear combinations can also be derived from this.
- $Q(\sqrt{x^{-1} - 1})$ is convex for $0 < x < 1$ (convexity in mean square error, which is required for performance evaluation and optimization of an MMSE equalizer) [9][12].

We list below a number of properties, which complement the properties above and, to the best of our knowledge, have not appeared in the communication/information-theoretic literature so far.

---

[10] Form this, however, it does not follow that the corresponding average SER is not convex (concave).

**Lemma 1:** $Q(x)$ and $[1-Q(x)]$ and are log-concave, i.e. $\log Q(x)$ and $\log[1-Q(x)]$ are concave.

**Proof:** follows from the integration theorem for log-concave densities [1][35][36] since the Gaussian noise density is log-concave (this can be verified by direct differentiation of $\log p_\xi(\mathbf{x})$).

**Lemma 2:** the second derivative $Q(x)''_x$ can be bounded as follows:

$$0 \leq Q''_x \leq (Q')^2 / Q, \; x \geq 0 \tag{29}$$

**Proof:** follows from Lemma 1 and the basic log-concavity inequality.

**Lemma 3:** $Q(\sqrt{\gamma})$ is log-convex in the SNR $\gamma > 0$.

**Proof:** consider $f(x) = \ln Q(\sqrt{x})$; one obtains,

$$f''_x = \frac{e^{-x/2}}{4\sqrt{2\pi}Q^2(\sqrt{x})} \frac{x+1}{x\sqrt{x}} \left( Q(\sqrt{x}) - \frac{\sqrt{x}e^{-x/2}}{\sqrt{2\pi}(1+x)} \right) \geq 0 \tag{30}$$

where the inequality follows from the known bound for $Q(x)$ [24],

$$Q(x) \geq \frac{xe^{-x^2/2}}{\sqrt{2\pi}(1+x^2)} \tag{31}$$

It should be pointed out that log-convexity is a stronger property than just convexity: while the later implies that $Q(\sqrt{\gamma})''_\gamma \geq 0$, the former implies that

$$Q(\sqrt{\gamma})''_\gamma \geq \frac{(Q(\sqrt{\gamma})'_\gamma)^2}{Q(\sqrt{\gamma})} = \frac{e^{-\gamma/2}}{2\pi Q(\sqrt{\gamma})} > 0 \tag{32}$$

Since $Q^2$ finds its way into some error rate expressions (see e.g. [13][16][17][32]), we list below its convexity properties.

**Lemma 4:** $Q(x)^2$, $[1-Q(x)]^2$ are log-concave, $Q(\sqrt{\gamma})^2$ is log-convex.

**Proof**: follows directly from Lemma 1.

**Lemma 5**: $Q(x)^2$ and $Q(\sqrt{x})^2$ are convex.

**Proof**: from the convexity of $Q(x)$, $Q(\sqrt{x})$ and using the composition rule [1].

## VII. APPLICATIONS

As it was emphasized above, convexity/concavity properties are important for optimization problems [1][2]. Below we consider some applications in digital communications, which include optimum power allocation for the ZF-SIC V-BLAST, optimum power/time sharing for the transmitter and jammer optimization, an implication for fading channels and an optimization of a unitary-precoded OFDM system.

## A. Optimum Power Allocation for the V-BLAST Algorithm

Consider the block error rate (BLER), i.e. the probability of at least one error at the detected transmit vector, of the ZF-SIC V-BLAST [3]-[5]:

$$P_B(\alpha_1...\alpha_m) = 1 - \prod_{i=1}^{m}(1 - P_e(\alpha_i\gamma_i)) \qquad (33)$$

where $P_e$ is the SER for the constellation in use, $\gamma_i$ is the SNR of $i$-th stream with uniform power allocation, $\alpha_i$ is the fraction of the total transmit power allocated to $i$-th stream (the uniform power allocation corresponds to $\alpha_i = 1$), $m$ is the number of streams (transmitters). Both instantaneous and average $P_e$ can be used in (33). Using the BLER as an objective, the following optimization problem can be formulated [3]-[5]:

$$\min_{\{\alpha_1...\alpha_m\}} P_B, \text{ subject to } \sum_{i=1}^{m}\alpha_i = m \qquad (34)$$

where the constraint insures that the total transmit power is fixed. The theorem below extends corresponding results in [3]-[5] derived for QAM modulation to a broad class of cases.

**Theorem 11:** The optimization problem in (34) has a unique global solution for either: (i) 1-D or 2-D constellations in terms of instantaneous or average (in Rayleigh, Rice and Nakagami-fading channels) BLER, or (ii) for $M$-D symmetric constellations, $M \geq 1$, in terms of instantaneous BLER, or (iii) any constellation at sufficiently high SNR.

**Proof:** note that the problem in (34) is equivalent to $\max_{\{\alpha_1...\alpha_m\}} \sum_{i=1}^{m} \log(1 - P_e(\alpha_i\gamma_i))$. If $P_e$ is convex, $(1 - P_e)$ and $\log(1 - P_e)$ are concave [1]. Thus, the objective function is concave and hence the problem has a unique solution. By Theorem 1 and 10, this holds for all 1-D or 2-D constellations in the AWGN channel, or Rayleigh, Rice, or Nakagami fading channels if the average BLER is used. For $M \geq 1$ and a symmetric constellation, the uniqueness in terms of instantaneous BLER follows from Corollary 3.1. For any constellation at sufficiently high SNR, the uniqueness follows from Theorem 2.

We note that Theorem 11 also applies to optimum power allocation in OFDM systems [41].

## B. Optimum Power/Time Sharing for a Jammer

Based on the concavity/convexity properties in Theorem 5, this section extends the corresponding results in [7] to arbitrary multi-dimensional constellations in the AWGN channel.

Considering $P_e$ as a function of $P_N$, one formulates the following jammer optimization problem using power/time sharing [7]:

$$\max_{\substack{n \\ \{\alpha_1...\alpha_n\} \\ \{P_{N1}...P_{Nn}\}}} \sum_{i=1}^{n} \alpha_i P_e(P_{Ni}), \text{ subject to } \sum_{i=1}^{n} \alpha_i = 1, \ \sum_{i=1}^{n} \alpha_i P_{Ni} = P_N \qquad (35)$$

where the jammer splits its transmission into $n$ sub-intervals, $\alpha_i$ being the fractional length of $i$-th sub-interval and $P_{Ni}$ is its noise (jammer) power, with the average noise power = $P_N$. The objective function in (35) is the SER over the whole transmission interval. An immediate conclusion from (35) is that if $P_e(P_N)$ is concave, the power/time sharing does not help, i.e. the best strategy is no sharing: $n=1$, $\alpha_1 = 1$, $P_{N1} = P_N$. This can be seem from the basic concavity inequality,

$$\sum_{i=1}^{n} \alpha_i P_e(P_{Ni}) \leq P_e\left(\sum_{i=1}^{n} \alpha_i P_{Ni}\right) = P_e(P_N) \qquad (36)$$

Theorem 5 ensures that the optimization is possible, i.e. $P_e$ can be increased by power/time sharing under the fixed average noise power, unless the noise power is large, in which case the best strategy is always "on". The optimum $n$ follows immediately from Caratheodory theorem [7][25]: $n \leq 2$, where $n=1$ corresponds to no sharing so that the only non-trivial solution is $n=2$, i.e. two power levels are enough to achieve the optimum. Let $\tilde{P}_e(P_N)$ denotes the maximum in (35), where "~" denotes optimality. Similarly to [7], it has simple characterization:

**Lemma 5**: $\tilde{P}_e(P_N)$ is concave.

**Proof:** by contradiction[11]. If it is not concave, one can apply the sharing in (35) again to increase it. But that is impossible as two consecutive sharings are equivalent to a single one and hence the second one does not help. Thus, $\tilde{P}_e(P_N)$ has to be concave, in which case second sharing does not help, as (36) indicates. *Q.E.D.*

It also follows that $\tilde{P}_e(P_N)$ is the smallest concave function that upper-bounds $P_e(P_N)$ [1][7][25]. This fact, however, seems to be immaterial for our problem as we try to maximize $P_e$ so larger functions are naturally welcome.

Before finding the optimal solution, we give a sub-optimal one, which is simpler to characterize. For clarity of exposition, we consider two cases, both of which rely on Theorem 5.

*Case I*: there is a single inflection point $P_0$, $P''_{e|P_N}(P_0) = 0$. From Theorem 5,

$$P''_{e|P_N} > 0 \text{ if } P_N < P_0, \ P''_{e|P_N} < 0 \text{ if } P_N > P_0 \qquad (37)$$

In this case, the sub-optimum sharing is as follows:

---

[11] The original proof in [7] relied on an elaborate argument. The contradiction-type proof given here is much simpler.

**Theorem 12**: The sub-optimum solution to (35) is to use the single power level (always "on") $P_{N1} = P_N$ if $P_N \geq P_0$, and "on-off" strategy with the on-interval $\alpha_1 = P_N / P_0$, $P_{N1} = P_0$ if $P_N < P_0$,

$$\{\alpha_i, P_{Ni}, i=1...n\} = \begin{cases} n=1, \ \alpha_1 = 1, \ P_{N1} = P_N & \text{if } P_N \geq P_0 \\ n=2, \ \alpha_1 = \dfrac{P_N}{P_0}, \ P_{N1} = P_0, P_{N2} = 0 & \text{if } P_N < P_0 \end{cases} \quad (38)$$

which achieves the following SER,

$$\tilde{P}_e(P_N) = \begin{cases} P_e(P_N), & P_N \geq P_0 \\ \dfrac{P_N}{P_0} P_e(P_0), & P_N < P_0 \end{cases} \quad (39)$$

**Proof**: it is straightforward to verify that (39) corresponds to the strategy in (38). Using (37) under the conditions of Theorem 5, it follows that $\tilde{P}_e(P_N) \geq P_e(P_N)$. Thus, (38) is indeed a better strategy than no sharing. Q.E.D.

Intuitive explanation for (38) is that one eliminates the convex part of $P_e(P_N)$ by time/power sharing and the concave part is left intact (no optimization is required there). Indeed, it can be verified that $\tilde{P}''_{e|P_N} = 0$ if $P_N < P_0$ and $\tilde{P}''_{e|P_N} < 0$ if $P_N > P_0$. The solution in (38) is not optimum since the first derivative of $\tilde{P}_e(P_N)$ is discontinuous at $P_N = P_0$ and $\tilde{P}''_{e|P_N}(P_0) = +\infty$ (unless $P'_{e|P_N}(P_0) = P_e(P_0)/P_0$, in which case (38) gives the optimum solution) so that $\tilde{P}_e(P_N)$ is not concave, which means that further optimization is possible.

*Case II*: there are multiple inflection points $P_{0k}$, $k=1...M$. Similar sub-optimal strategy can be used in the case of multiple inflection points.

**Theorem 13**: the sub-optimal time/power sharing strategy of the jammer in this case is as follows,

$$\{\alpha_i, P_{Ni}, i=1...n\} = \begin{cases} n=1, \ \alpha_1 = 1, \ P_{N1} = P_N \text{ if } P_N \in D^- \\ n=2, \ \alpha_1 = \dfrac{P_{0(k+1)} - P_N}{P_{0(k+1)} - P_{0k}}, \\ P_{N1} = P_{0k}, P_{N2} = P_{0(k+1)} \end{cases} \text{ if } P_N \in D_k^+ \quad (40)$$

where $D_k^+ = [P_{0(k+1)}, P_{0k}]$ is k-th interval of convexity of $P_e(P_N)$, and $D^-$ denotes all concave intervals. This achieves the following SER,

$$\tilde{P}_e(P_N) = \begin{cases} P_e(P_N), & P_N \in D^- \\ \alpha_1 P_e(P_{0k}) + \alpha_2 P_e(P_{0(k+1)}), & P_N \in D_k^+ \end{cases} \quad (41)$$

**Proof**: similar to that of Theorem 12.

We now consider the optimal solution for the case of a single inflection point.

**Theorem 14**: the optimum power/time sharing strategy of the jammer for the case of a single inflection point is always "on" at high noise power, and "on-off" at low,

$$\{\alpha_i, P_{Ni}, i=1...n\} = \begin{cases} n=1, \ \alpha_1 = 1, \ P_{N1} = P_N & \text{if } P_N \geq P^* \\ n=2, \ \alpha_1 = \dfrac{P_N}{P^*}, \ P_{N1} = P^*, \ P_{N2} = 0 & \text{if } P_N < P^* \end{cases} \quad (42)$$

where $P^* \geq P_0$ is such that $P'_{e|P_N}(P^*) = P_e(P^*)/P^*$, which achieves the following SER,

$$\tilde{P}_e(P_N) = \begin{cases} P_e(P_N), & P_N \geq P^* \\ \dfrac{P_N}{P^*} P_e(P^*), & P_N < P^* \end{cases} \quad (43)$$

**Proof:** based the concavity/convexity properties in Theorem 5, follows along the same lines as that in [[7], Theorem 3].

Note that (42) is identical to (38) with the differently-defined threshold $P^*$, i.e. the optimum strategy can be obtained from the sub-optimum one by proper adjustment of the threshold. The intuition behind the optimum solution is almost the same as that of the sub-optimum one. The only difference is that the power/time sharing strategy extends into the concave part to ensure the continuity of the first derivative of $\tilde{P}_e(P_N)$ so that its second derivative is always non-positive and hence $\tilde{P}_e(P_N)$ is concave and no further optimization is possible by time/power sharing. The optimal solution for the case of multiple inflection points can be constructed in a similar way.

*C. Optimum Time/Power Sharing for the Transmitter*

Similarly to the jammer problem above, the optimization problem can be formulated for the transmitter, with the objective to reduce the SER. In fact, these two problems are equivalent, via the substitutions,

$$P_c \to P_e, \ \gamma \to P_N \quad (44)$$

For completeness, we formulate below the main results.

**Theorem 15**:

a) If $P_c(\gamma)$ is concave, e.g. for 1-D and 2-D constellations, the optimum transmission strategy is always "on", without sharing (i.e. power/time sharing does not help to reduce the SER, which was the case in [7] for binary modulations).

b) If $P_c(\gamma)$ is not concave, e.g. for some $M$-D constellations, $M \geq 3$, (i) the sub-optimum transmitter strategy is given by Theorems 12 and 13, and (ii) the optimum transmitter strategy is given by Theorem 14, both with the substitutions in (44).

Comparing these results to those in the previous section, we conclude that the jammer is in better position compared to the transmitter for 1-D and 2-D constellations, as the former can use power/time sharing to optimize its transmission strategy while the latter can not.

*D. Implication for Fading Channels*

The convexity property in Theorem 1 has a profound implication for arbitrary-fading channels. The following result is a straightforward consequence of the basic convexity inequality (Jensen inequality) and the result in Theorem 1.

**Theorem 16**: If $P_e(\gamma)$ is convex in the non-fading AWGN channel, e.g for 1-D and 2-D constellations, fading never reduces the SER (compared to the non-fading channel at the same average SNR), i.e. "*fading is never good in low dimensions*",

$$\overline{P_e(\gamma)} \geq P_e(\bar{\gamma}) \tag{45}$$

where $\bar{x}$ denotes mean value of $x$.

Based on Theorem 2, this result also extends to higher-D constellations at high SNR; at small SNR, there are "good" types of fading, which reduce the SER. Intuitively, this corresponds to the optimum (or sub-optimum) transmitter strategy of the previous section, since the time/power sharing strategy of the transmitter can be considered as "fading".

*E. Unitary Precoding in OFDM Systems*

Let us now apply the convexity/concavity results to bound error rates of precoded OFDM systems, which also reveals what is the best and worst precoding. Following [9], we consider an OFDM system with a unitary precoder (i.e. the Tx precoding matrix $\mathbf{T}$ is unitary); conventional OFDM system (without precoding) corresponds to $\mathbf{T} = \mathbf{I}$, where $\mathbf{I}$ is the identity matrix; the single-carrier cyclic prefix (SC-CP) system, with equalization done at the receiver using Fast Fourier Transform (FFT) and inverse FFT (IFFT), corresponds to $\mathbf{T} = \mathbf{W}$, where $\mathbf{W}$ is the FFT matrix. Based of the convexity/concavity property of $Q(1/\sqrt{x})$, it has been shown in [9] that the error rate $P_e^T$ of the unitary-precoded OFDM system with QPSK modulation and arbitrary $\mathbf{T}$ can be bounded as,

$$P_e^{OFDM} \leq P_e^T \leq P_e^{SC-CP}, \text{ at low SNR}$$
$$P_e^{OFDM} \geq P_e^T \geq P_e^{SC-CP}, \text{ at high SNR} \quad (46)$$

where $P_e^{OFDM}$ and $P_e^{SC-CP}$ are the error rates of the conventional OFDM (no precoding) and the SC-CP system with the ZF equalizer, respectively, and high/low SNR regions are quantified in [9]. It follows form (46) that the SC-CP system ($\mathbf{T} = \mathbf{W}$) is the best unitary precoding at high SNR, and the conventional OFDM ($\mathbf{T} = \mathbf{I}$) is the best at low SNR[12].

Using Theorem 5, the result in (46) immediately applies to arbitrary multidimensional constellations (with low and high SNR regions defined based on the thresholds in Theorem 5 and corollary 5.1), which may also include coding, i.e. it is a consequence of the transmission strategy rather than a particular constellation used, where the latter determines only the low and high SNR thresholds.

## VIII. CONCLUSIONS

Convexity/concavity properties of error rates of the ML detector in non-fading and fading AWGN channels in terms of SNR and noise power have been studied. It has been shown that the SER is always convex in SNR for 1-D and 2-D constellations, but may be non-convex in higher dimensions at low to intermediate SNR, being always convex at high SNR. The pairwise error probability and also the BER are convex at high SNR. The SER is concave in noise power at low SNR (large noise power), convex at high SNR (small noise power) and has an odd number of inflection points in-between. Universal bounds on the SER 1st and 2nd derivatives have been derived, which are the functions of SNR and constellation dimensionality only and are independent of the constellation geometry. A number of related properties of the Q-function have been discussed. The applications of these results to optimization problems were presented, which included the optimum power allocation in the spatial multiplexing system (ZF-SIC V-BLAST), optimum power/time sharing strategy for transmitter and jammer, optimal orthogonal precoding for OFDM systems and implication for fading channels. These results extend to the interference channel when the Gaussian approximation of interference is used, and also to the BER when it can be expressed as a non-negative linear combination of individual symbol error rates, or when it can be approximated as $\text{BER} \approx \text{SER} / \log_2 M$. The BER is always convex at high SNR. Furthermore, the reported results also apply to coded systems, by considering codewords as symbols of an extended multi-dimensional constellation.

---

[12] As a side remark, we note that the best and worst precoding here follow immediately from the basic concavity/convexity inequalities, without the complications of explicitly solving an optimization problem (e.g. via Lagrange multipliers), which emphasizes once more the importance of convexity/concavity properties.

The convexity/concavity properties of error rates studied above open up a possibility to apply the convex optimization techniques to many problems in digital communications in a systematic and rigorous way, thus providing a missing generic link between digital communications and convex optimization. Furthermore, generic convexity/concavity properties provide significant insights into constellation-independent system properties, i.e. those that depend on system configuration and transmission strategy (e.g. V-BLAST, power/time sharing, OFDM precoding) rather than a particular constellation in use. Optimum or near-optimum transmission strategies and their performance can sometimes be derived directly based on the basic convexity/concavity inequalities, without complex machinery of analytical or numerical optimization.

IX. ACKNOWLEDGEMENTS

The first author would like to thank F. Kschischang and M. Dohler for fruitful discussions, and D. Palomar for his interest in this work.

## XI. APPENDIX A: PROOFS

**Proof of Theorem 1**: consider first $P''_{ci|\gamma}$, which can be expressed as

$$P''_{ci|\gamma} = \int_{\Omega_i} \frac{d^2 p_\xi(\mathbf{x})}{d\gamma^2} d\mathbf{x} \tag{A1}$$

where we have exchanged the integral and derivative since the integration boundary does not depend on $\gamma$ and $p_\xi(\mathbf{x})$ is continuous in $\mathbf{x}$, $\gamma$, and is differentiable in $\gamma$. After some manipulations, the derivative in (A1) can be presented in the following form,

$$\frac{d^2 p_\xi(\mathbf{x})}{d\gamma^2} = \frac{1}{4}\left(\frac{\gamma}{2\pi}\right)^{\frac{n}{2}} e^{-\frac{\gamma|\mathbf{x}|^2}{2}} f\left(|\mathbf{x}|^2\right)$$
$$f(t) = \left(t - \frac{\alpha_1}{\gamma}\right)\left(t - \frac{\alpha_2}{\gamma}\right), \quad \alpha_1 = n + \sqrt{2n}, \quad \alpha_2 = n - \sqrt{2n} \tag{A2}$$

where $\alpha_1 > 0$, $\alpha_2 \leq 0$, so that $f\left(|\mathbf{x}|^2\right) \leq 0$ if $|\mathbf{x}|^2 \leq \alpha_1/\gamma$, and $f\left(|\mathbf{x}|^2\right) > 0$ otherwise. Consider three different cases.

*Case 1:*

$$d^2_{\max,i} \leq \frac{\alpha_1}{\gamma} \tag{A3}$$

where $d_{\max,i}$ is the maximum distance from the origin to the boundary of $\Omega_i$. In this case, $f\left(|\mathbf{x}|^2\right) \leq 0$ $\forall \mathbf{x} \in \Omega_i$ so that the integral in (A1) is clearly negative and (8) follows. Fig. 1 illustrates this case. This is a small-SNR mode since (A3) implies that $\gamma \leq \alpha_1 / d_{\max,i}^2$. In the same way, one can prove Theorem 4(b).

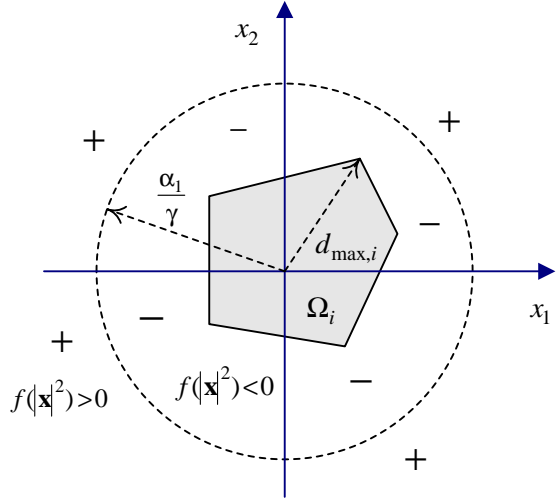

Fig. 1. Two-dimensional illustration of the problem geometry for Case 1. The decision region $\Omega_i$ is shaded. $f\left(|\mathbf{x}|^2\right)$ has a sign as indicated by "+" and "-".

*Case 2:*

$$d_{\min,i}^2 \geq \frac{\alpha_1}{\gamma} \tag{A4}$$

where $d_{\min,i} = \min_j (b_j)$ is the minimum distance from the origin to the boundary of $\Omega_i$. Fig. 2 illustrates this case. This is a large-SNR mode since (14) implies that $\gamma \geq \alpha_1 / d_{\min,i}^2$. In this case, $f\left(|\mathbf{x}|^2\right) \geq 0$ $\forall \mathbf{x} \in \left(\mathbf{R}^n - \Omega_i\right)$, where $\mathbf{R}^n$ is the n-dimensional space, and the difference of two sets $S_1$ and $S_2$ is defined as $\left(S_1 - S_2\right) = \left\{\mathbf{x} | \mathbf{x} \in S_1, \mathbf{x} \notin S_2\right\}$, so that $\left(\mathbf{R}^n - \Omega_i\right) = \left\{\mathbf{x} | \mathbf{x} \notin \Omega_i\right\}$ is the complement of $\Omega_i$. The integral in (A1) can be upper bounded as

$$P''_{ci|\gamma} = \int_{\Omega_i} \frac{d^2 p_\xi(\mathbf{x})}{d\gamma^2} d\mathbf{x} < \int_{\mathbf{R}^n} \frac{d^2 p_\xi(\mathbf{x})}{d\gamma^2} d\mathbf{x} = 0 \tag{A5}$$

where we have used the fact that $\int_{\mathbf{R}^n} p_\xi(\mathbf{x}) d\mathbf{x} = 1$. In this case, the pairwise error probability is also convex (see Theorem 4).

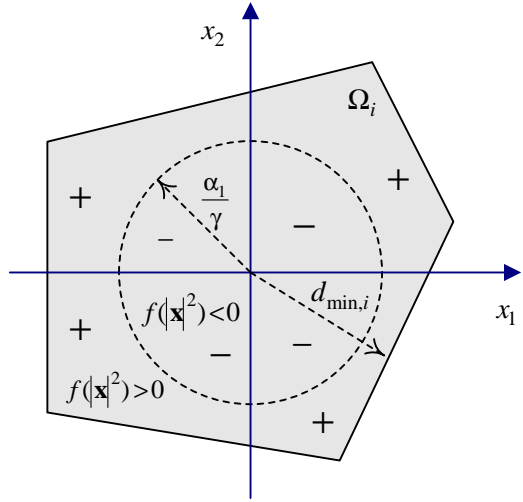

Fig. 2. Two-dimensional illustration of the problem geometry for Case 2.

*Case 3:*

$$d_{\min,i}^2 < \frac{\alpha_1}{\gamma} < d_{\max,i}^2 \tag{A6}$$

This is an intermediate-SNR mode since (A6) implies that $\alpha_1 / d_{\max,i}^2 \leq \gamma \leq \alpha_1 / d_{\min,i}^2$. Fig. 3 illustrates this case. Separating the decision region $\Omega_i$ into two sub-regions, $\Omega_i = \Omega_a + \Omega_b$, $\Omega_a = \Omega_i - \Omega_i \cap \Omega_{con}$, $\Omega_b = \Omega_i \cap \Omega_{con}$, where $\Omega_{con}$ is (are) the cone(s) whose base(s) is (are) the intersection(s) of the planes $\mathbf{a}_j^T \mathbf{x} = b_j$ (boundaries of the decision region $\Omega_i$) and the ball $|\mathbf{x}|^2 \leq \alpha_1 / \gamma$; the vertex of the cone(s) is the origin $\mathbf{x} = 0$. Clearly,

$$\int_{\Omega_b} \frac{d^2 p_\xi(\mathbf{x})}{d\gamma^2} d\mathbf{x} \leq 0 \tag{A7}$$

The integral over $\Omega_a$ can be bounded as

$$\int_{\Omega_a} \frac{d^2 p_\xi(\mathbf{x})}{d\gamma^2} d\mathbf{x} < \int_{(\mathbf{R}^n - \Omega_{con})} \frac{d^2 p_\xi(\mathbf{x})}{d\gamma^2} d\mathbf{x} = 0, \tag{A8}$$

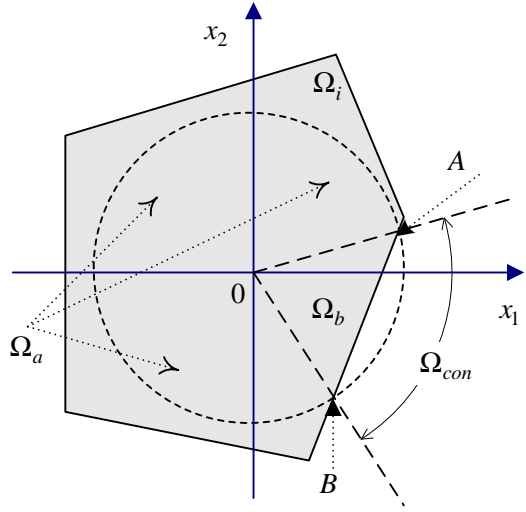

Fig. 3. Two-dimentional illustration of the problem geometry for Case 3. The cone $\Omega_{con}$ is build on the OA and OB rays. $\Omega_b$ is the triangle AOB.

where the inequality follows from the fact that $f(|\mathbf{x}|^2) > 0 \ \forall \mathbf{x} \in (\mathbf{R}^n - \Omega_{con} \cup \Omega_i)$, and the equality follows from the fact that

$$\int_{(\mathbf{R}^n - \Omega_{con})} p_\xi(\mathbf{x}) d\mathbf{x} = \frac{\Psi_{tot} - \Psi_{con}}{\Psi_{tot}} \tag{A9}$$

where $\Psi_{con}$ is the fixed solid angle spanned by $\Omega_{con}$ and $\Psi_{tot}$ is the total solid angle in $\mathbf{R}^n$, both of them are independent of $\gamma$ [13]. Note that (A9) follows from the spherical symmetry of $p_\xi(\mathbf{x})$ (see (2)). Combining (A8) and (A7), one obtains $P''_{ci|\gamma} < 0$.

Thus, $P_{ci}$ is concave and, hence, $P_{ei}$ is convex in all three cases. Using the fact that a non-negative weighted sum of concave (convex) functions is concave (convex) [1], one concludes that $P_c$ is also concave and hence $P_e$ is convex. *Q.E.D.*

**Proof of Theorem 2**: First, we note that for $n > 2$, $\alpha_1 > \alpha_2 > 0$. In the large SNR case (9), $d^2_{min,i} \geq \alpha_1/\gamma$ so that $f(|\mathbf{x}|^2) > 0 \ \forall \mathbf{x} \in (\mathbf{R}^n - \Omega_i)$, and the integral for $P_{ci}$ can be upper bounded as in (A5) from which 2.1. follows. In the small SNR mode, $d^2_{max,i} \leq \alpha_2/\gamma < \alpha_1/\gamma$ so that $f(|\mathbf{x}|^2) \geq 0 \ \forall \mathbf{x} \in \Omega_i$ and the integral in (A1) is positive. Since $P_{ci}$ is concave in the large SNR mode, $P''_{ci|\gamma} < 0$, and is convex in the small SNR mode, $P''_{ci|\gamma} > 0$, there must be an odd number of inflection points, $P''_{ci|\gamma} = 0$, in-between (by the continuity argument applied to $P''_{ci|\gamma}$). *Q.E.D.*

---

[13] $\Omega_{con}$ may be an intersection of several cones, in which case $\Psi_{con}$ is the total solid angle spanned by this intersection, which is still independent of $\gamma$

**Proof of Theorem 5**: Similarly to (A2), $d^2 p_\xi(\mathbf{x})/dP_N^2$ can be expressed as

$$\frac{d^2 p_\xi(\mathbf{x})}{dP_N^2} = \frac{1}{4P_N^4}\left(\frac{1}{2\pi P_N}\right)^{\frac{n}{2}} e^{-\frac{|\mathbf{x}|^2}{2P_N}} f^*\left(|\mathbf{x}|^2\right)$$

$$f^*\left(|\mathbf{x}|^2\right) = \left(|\mathbf{x}|^2 - \beta_1 P_N\right)\left(|\mathbf{x}|^2 - \beta_2 P_N\right), \tag{A10}$$

$$\beta_1 = n+2+\sqrt{2(n+2)}, \quad \beta_2 = n+2-\sqrt{2(n+2)}$$

where $\beta_1 > \beta_2 > 0$. Using (A10) in the proof of Theorem 2, Theorem 5 follows. *Q.E.D.*

**Proof of Theorem 6:** the derivative $dp_\xi(\mathbf{x})/d\gamma$ can be expressed as

$$\frac{dp_\xi(\mathbf{x})}{d\gamma} = \frac{1}{2}\left(\frac{\gamma}{2\pi}\right)^{\frac{n}{2}}\left(\frac{n}{\gamma} - |\mathbf{x}|^2\right)e^{-\frac{\gamma|\mathbf{x}|^2}{2}} \tag{A11}$$

Noting that $dp_\xi(\mathbf{x})/d\gamma \geq 0$ if and only if $|\mathbf{x}|^2 \leq n/\gamma$, $P'_{ci|\gamma}$ can be upper-bounded as

$$P'_{ci|\gamma} = \int_{\Omega_i} \frac{dp_\xi(\mathbf{x})}{d\gamma} d\mathbf{x} \leq \int_{C^+} \frac{dp_\xi(\mathbf{x})}{d\gamma} d\mathbf{x} \tag{A12}$$

where $C^+$ is the ball of radius $\sqrt{n/\gamma}$, $C^+ = \{\mathbf{x} | |\mathbf{x}|^2 \leq n/\gamma\}$. The last integral in (A12) can be evaluated in a closed form by using the spherical coordinates and relying on the spherical symmetry of $p_\xi(\mathbf{x})$ [22]. Specifically, the integral of $p_\xi(\mathbf{x})$ over the sphere of radius $R$ is

$$\int_{|\mathbf{x}|=R} p_\xi(\mathbf{x})d\mathbf{x} = \frac{1}{\Gamma(n/2)}\gamma^*\left(\frac{n}{2};\frac{\gamma}{2}R^2\right) \tag{A13}$$

where $\gamma^*(x;y)$ is the incomplete gamma-function [23],

$$\gamma^*(x;y) = \int_0^y e^{-t} t^{x-1} dt \tag{A14}$$

and $\Gamma(x) = \gamma^*(x;\infty)$ is the complete gamma-function. Using (A13), one obtains

$$\int_{C^+} \frac{dp_\xi(\mathbf{x})}{d\gamma} d\mathbf{x} = \frac{(n/2)^{n/2} e^{-n/2}}{\gamma \Gamma(n/2)} \tag{A15}$$

Since (A15) is independent of $\Omega_i$, it also applies to $P'_{ci|\gamma}$. This proves the lower bound in Theorem 6. The upper bound is obvious; its formal proof can be obtained along the lines of that of Theorem 1. *Q.E.D.*

Sergey Loyka (M'96–SM'04) was born in Minsk, Belarus. He received the Ph.D. degree in Radio Engineering from the Belorussian State University of Informatics and Radioelectronics, Minsk, Belarus in 1995 and the M.S. degree with honors from Minsk Radioengineering Institute, Minsk, Belarus in 1992. Since 2001 he has been a faculty member at the School of Information Technology and Engineering, University of Ottawa, Canada. Prior to that, he was a research fellow in the Laboratory of Communications and Integrated Microelectronics (LACIME) of Ecole de Technologie Superieure, Montreal, Canada; a senior scientist at the Electromagnetic Compatibility Laboratory, Belorussian State University of Informatics and Radioelectronics, Minsk , Belarus; an invited scientist at the Laboratory of Electromagnetism and Acoustic, Swiss Federal Institute of Technology – Lausanne (EPFL), Switzerland. His research areas include wireless communications and networks, MIMO systems and smart antennas, RF system modeling and simulation, and electromagnetic compatibility, in which he has published extensively.

Victoria Kostina received a Bachelor's degree with honors in applied mathematics and physics from the Moscow Institute of Physics and Technology, Russia, in 2004, and a Master's degree in electrical engineering from the University of Ottawa, Canada, in 2006. She is currently working towards her Ph.D. at Princeton University, USA. Her research interests are in information theory, coding and wireless communications.

Francois Gagnon (S'87-M'87-SM'99) was born in Québec City, PQ., Canada. He received the B.Eng. and Ph.D. degrees in electrical engineering from École Polytechnique de Montréal, Montréal, P.Q., Canada. Since 1991, he has been a Professor in the Department of Electrical Engineering, École de technologie supérieure, Montréal, Canada. He has chaired the department from 1999 to 2001 and is now the holder of the Ultra Electronics (TCS) Chair in Wireless Communication in the same university. His research interest covers wireless high speed communications, modulation, coding, high speed DSP implementations and military point to point communications. He has been very involved in the creation of the new generation of High Capacity Line Of Sight military radios offered by the Canadian Marconi Corporation, which is now Ultra Electronics Tactical Communication Systems. The company has received, for this product, a 'Coin of excellence' from the US army for performance and reliability.